# Porous Alumina Based Capacitive MEMS RH Sensor

László Juhász[1], András Vass-Várnai[1], Veronika Timár-Horváth[1], Marc P.Y. Desmulliez[2], Resh S. Dhariwal[2]

[1]Budapest University of Technology and Economics (BME), Dept. of Electron Devices, Budapest, Hungary
[2]MicroSystems Engineering Centre (MISEC), Heriot-Watt University (HWU), Edinburgh, United Kingdom

The aim of a joint research and development project at the BME and HWU is to produce a cheap, reliable, low-power and CMOS-MEMS process compatible capacitive type relative humidity (RH) sensor that can be incorporated into a state-of-the-art, wireless sensor network. In this paper we discuss the preparation of our new capacitive structure based on post-CMOS MEMS processes and the methods which were used to characterize the thin film porous alumina sensing layer. The average sensitivity is approx. 15 pF/RH% which is more than a magnitude higher than the values found in the literature. The sensor is equipped with integrated resistive heating, which can be used for maintenance to reduce drift, or for keeping the sensing layer at elevated temperature, as an alternative method for temperature-dependence cancellation.

## I. INTRODUCTION

Nowadays, humidity is an important factor to consider in certain industrial and agricultural processes, in the preservation of objects of our art heritage and in other numerous areas of our lives. There are numerous old and new methods and devices to measure quantities that are related to the humidity, namely the absolute humidity, the relative humidity (RH) and the dew point [1].

Capacitive sensors are based on the changes of the dielectric property of thick or thin films upon water vapor uptake which depends on the relative humidity content of the surrounding media. Due to the polar structure of the $H_2O$ molecule, water exhibits a very high permittivity ($\kappa_w$=80) at room temperature. The permittivity of dielectric films shows a huge increase upon adsorption of water. In a porous dielectric the air in the voids is replaced by adsorbed vapor as the ambient humidity level increases therefore it is worth using porous dielectrics instead of compact layers.

Capacitive sensors – as well as other absorption-based humidity sensors – typically show a non-linear behavior as function of RH. This behavior can be described by the following phenomenological equation:

$$\frac{C_w}{C_d} = \left(\frac{\kappa_w}{\kappa_d}\right)^n \quad (1)$$

where $\kappa_d$ and $\kappa_w$, are the permittivity coefficients of the dielectric material at the dry and wet states, respectively, $C_w$ and $C_w$ are the corresponding capacitances and $n$ is a factor related to (the morphology of) the dielectric film [1].

Because of the immense surface-to-volume ratio and the abundant void fraction, very high sensitivities can be obtained with porous ceramics. One of the ceramics to be used is porous $Al_2O_3$ which has proven to be stable at elevated temperatures and at high humidity levels. A well established method for porous $Al_2O_3$ preparation is the electrochemical oxidation of Al films under anodic bias. It was demonstrated in [2], that it is possible to produce highly sensitive, micro-sized AAO (anodic aluminum oxide) sensors on silicon chips using thin film aluminum as starting material.

The quality of the $Al_2O_3$ layers and consequently the sensor behavior (sensitivity, humidity range, response time, stability etc.) strongly depend on the thickness of the porous layer and the density and size of the pores [2] [3]. The structure of porous alumina prepared by the aforementioned method can be influenced by the technological parameters such as concentration and temperature of the electrolyte and the current density (or voltage) in the anodizing cell [2] [4] [5] [6].

Theoretically, the best solution to cover the total humidity range (0–100%) is to produce all size of pores: micro-, meso- and macropores[1]. According to a recent study [5] a highly reproducible, wide-range humidity sensor can be achieved using nanodimensional pores of a narrow size distribution.

The response time of ceramic humidity sensors is generally limited by diffusion [7]. Moreover, ceramics are highly sensitive to contaminants such as dust and smoke. These sensors require maintenance which consists in having the condensed vapor evaporated by heating up them from time to time. The temporary heating could also be a solution to reduce the drift due to the formation of chemisorbed $OH^-$ groups [5] [8].

## II. EXPERIMENTAL

Boron doped (p-type) 2-inch silicon wafer was used as substrate. Wafers were cleaned in RBS detergent, refreshed in dilute (1:20) HF, boiled in dionised water and, after drying, oxidized in 100 l/hour dry $O_2$ flow at 1100 °C for 40 minutes, resulting a 100 nm thick $SiO_2$ layer serving as insulation [2] [9] and as adhesive layer [2]. E-beam vacuum vapor deposition was carried out to create a titanium (100 nm) and aluminum (99.9% purity, 500 nm) thin film layers structure. The titanium served as the lower electrode of the vertical capacitive structure. Using the first mask for photolithography (with AZ-type thin resist) the aluminum was selectively anodized in aqueous solution of 10 wt% sulfuric acid [5] opposed to the previous experiments [2] where 23.1 wt% solution had been used. The anodic current density was kept at a constant value (8 mA/cm$^2$) [4]. After photoresist removal the remaining aluminum was etched away using a selective wet etchant (16 H$_3$PO$_4$ : 1 HNO$_3$). Titanium and copper were evaporated on the top of the wafer and windows were opened in thick AZ-type photoresist to enable the formation of the upper grid-electrode and contact surfaces using gold electroplating. After stripping the photoresist, wet copper etchant (10% Na$_2$S$_2$O$_8$) was used. The third photolithographic step protected the meanders of the resistive heating and the surrounding surfaces of the contacts during the titanium wet etching (10 NH$_4$OH: 1 H$_2$O$_2$). The ready-made wafer was cut to dices and the selected dices were fixed onto TO-type headers with silicon-based glue. The gold contact pads of each chip were connected to the leads with thermocompression using 50 μm gold wire. The completed capacitive sensors structure can be seen on Fig. 1.

---

[1] Micropores have widths smaller than 2 nm. Mesopores have widths between 2 and 50 nm. Macropores have widths larger than 50 nm. [10]

During the selective anodization of the aluminum layer two additional effects have to be taken into consideration. One is the distortion of the planned structure due to under-etching on the perimeter of the photoresist windows; the other is the possible formation of small dots on seemingly random sites due to inadequate resist thickness or quality.

The packaged sensors were tested in controlled humidity environment to acquire RH–capacitance characteristics, but additional methods were also applied to examine the previously prepared porous alumina layers to explain the results.

### III. RESULTS AND DISCUSSION

To test the devices, two types of environment tests were used: fixed-point humidity environment over five different saturated salt solutions at 25 °C [11], and a humidity chamber (ESPEC SH-241). Capacitance measurements were performed using HAMEG LC meter at 16 kHz. Typical sensing characteristics are shown in Fig. 2.

Both the characteristics measured over saturated salt solutions and in the humidity chamber show the same sensitivity of our structure and match the previous results obtained with another test-structure [2], with the exception of the sensitivity. The average sensitivity is three times higher: approx. 15 pF/RH%. This is much better than the typical 0.2-0.5 pF/RH% as found in [12].

The sensing characteristics are non-linear, as desired, both a large hysteresis loop (Fig. 3.) and an enhancement in the sensitivity over approx. 80% RH can be observed. The characteristics are also temperature-dependent. In the next sections these properties are investigated.

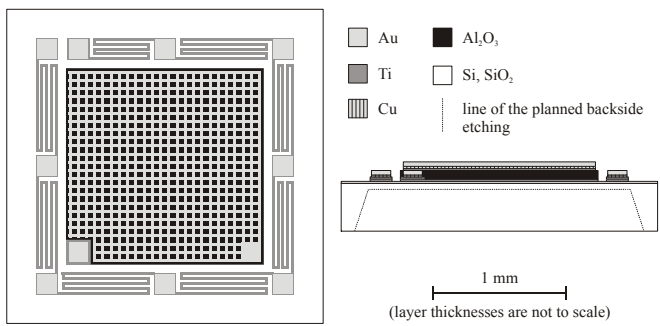

Fig. 1. Top view and cross-section of the capacitive RH sensor.

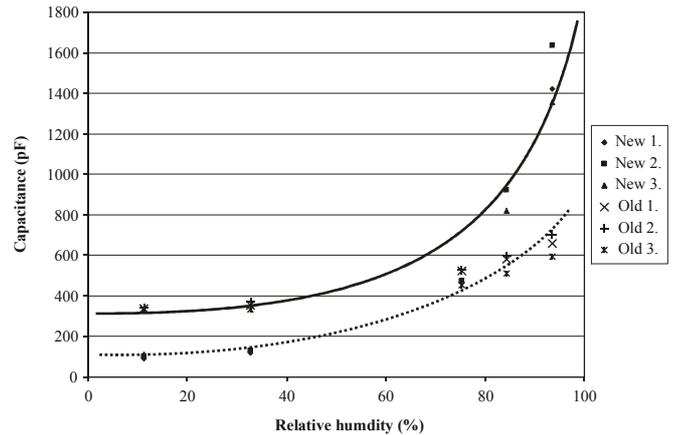

Fig. 2. Typical sensing characteristics of the sensors compared to the old test-structure.

*A. Hysteresis loop and linearity*

At relatively low RH levels very thin layers develop from water molecules on the pore walls and heat of condensation releases. This phenomenon can be explained by the BET theory of adsorption on microporous materials [13]. The BET theory assumes that the solid material has fixed number of adsorption sites and the adsorbed layers can be several molecules thick. It is also put forward, that the heat of adsorption in all layers beyond the first is equal to the latent heat of condensation:

$$\frac{p}{v(p_0 - p)} = \frac{1}{v_m c} + \frac{c-1}{v_m c} \frac{p}{p_0} \qquad (2)$$

where $p$ is pressure of the gas, $p_0$ is the saturated vapor pressure, $v$ is the quantity of the adsorbed gas, $v_m$ is the monolayer capacity of the surface (the quantity of one adsorbed layer), $c$ is a temperature-dependent value:

$$c = e^{\frac{E_1 - E_L}{RT}} \qquad (3)$$

where $E_1$ is the adsorption heat of the first layer and $E_L$ is the heat of condensation. A further assumption is that the vapor condenses on the adsorbent as a liquid when its pressure reaches the saturated vapor pressure, i.e., at $p=p_0$ the number of the adsorbed layers is infinite.

An equation similar to (2) might describe the adsorption branch of the response curve shown in Fig. 3.

As the humidity level increases, the pores start to fill. This filling can be explained by the capillary condensation theory. In case of higher RH it has been shown by Kelvin that, at a given pressure, vapor condenses into those pores, which have radii smaller than the Kelvin radius, given by following equation:

$$\ln(p/p_0) = -\frac{2\gamma V}{rRT} \cdot \cos\theta \qquad (4)$$

where $\gamma$ is the surface tension of the liquid, $V$ is the molecular volume and $R$ and $T$ have their usual meanings. The negative sign implies for $\theta<90°$, i.e. $p$ is less than $p_0$.

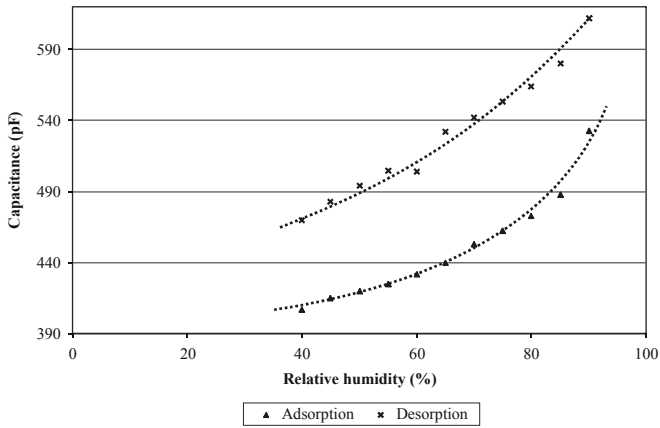

Fig. 3. Adsorption and desorption branches measured on one of the test-structures.

The capillary condensation might cause hysteresis in the characteristics [13]. As *p* increases, wider and wider pores will be filled, as *p* decreases, the pores get empty. There are two different contact angles on the two branches of the loop, while the pores are filling, there is an advancing contact angle, $\theta_A$, and while they are emptying, there is a "receding" contact angle, $\theta_R$. Since $\theta_A$ is greater than $\theta_R$, according to the equation (4), the value of *p* for a given value of *r* is less during desorption than adsorption causing the hysteresis (Fig. 3.). This gives the explanation of the upper part of the sensing curve (desorption). Over 80% RH level physical adsorption will be taken over by capillary condensation in the whole porous structure.

Wetting measurements were conducted on the alumina surface formed on the oxidized silicon wafer to measure the different contact-angles. We have approximated the surface as one side of an infinitely wide pore. The contact-angles were measured by the goniometric method using an OCA 40 Automatic Contact Angle Meter. Small droplets of water (8 µl) were dropped onto the surface. The advancing contact angles were approx. 70°, the receding 38°. The latter was taken up with the continuous withdrawal of the droplet. The difference between the advancing and receding contact angles might be the main reason of the hysteresis.

*B. Temperature-dependence of the sensor*

The other major side effect is the temperature-dependence of the vapor adsorption phenomena in the porous structure, which has been proven by measurements. A constant RH was maintained in the humidity chamber and the temperature was varied in the range of 5–95 °C. Prior to the measurements the sensors were heated up to 100 °C and kept at this temperature for 10 minutes resulting in dry pores. The results are shown in Fig. 4.

A possible explanation of this phenomenon was given by R. K. Nahar [7]. He suggested that, besides the capillary condensation of water vapor, there is lateral moisture diffusion into the pore walls as well, causing a virtual widening of pores especially at higher temperatures, following the exponential diffusion rules. The T–C curve's hysteresis might be caused by the difference between the diffusion mechanism of the water molecules penetrating into or exiting from the structure of the pore walls or the grain boundaries of the alumina layer. This phenomenon might affect the original sensing mechanism as well.

The resulting characteristics (type IV isotherms), showing a hysteresis loop are typical in case of adsorption on mesoporous adsorbents [13]. To obtain some information of the alumina layer, we have made sorption measurements, following the IUPAC (International Union of Pure and Applied Chemistry) recommendations [10].

A Hydrosorb TM 100 computer controlled volumetric instrument (Quantachrome, USA) chamber was used for the purpose of sorption measurements. Unfortunately the 2" wafers with the alumina coating have significantly lower surface area than the required 1.5 m$^2$. Our only choice was modeling the porous structure by some alumina powder.

Prior to the measurements, the same powder was treated at 20 °C and 200 °C in low vacuum (2.7 Pa) for two days. Both measurements were performed at the constant temperature of 20 °C. The resulting isotherms are shown in Fig. 5.

Both curves can be considered type IV isotherms according to IUPAC classification. According to the results, it is obvious that the heat treatment played an important role. The powder could take up significantly more water after it was outgased at 200 °C than at 20 °C. There is an inflexion at approx. 15% RH ($p/p_0$=0.15) in both isotherms, showing the formation of a monomolecular layer of water molecules on the porous structure. From this point up to 70% RH ($p/p_0$=0.70) the isotherms can be regarded linear. Over 70%, due to the capillary condensation, the rate of adsorption increases in the case of both isotherms. This behavior is similar to the RH–capacitance characteristics of the sensor.

The fact that after the powder was outgased at 200 °C and the hysteresis did not close in the entire RH region, implies that, beside the physical adsorption, chemisorption can occur. Their different activation energy might explain the sensors temperature-dependence.

*C. Integrated heating*

Humidity sensors based on porous ceramics like alumina require maintenance from time to time, which can be assured by heating up the sensing structure. This solution reduces the drift caused by the formation of chemisorbed OH$^-$ groups.

The integrated heating elements could also be used to keep the sensing layer at elevated temperature [8], which could be another solution for treating the temperature-dependence besides the electronic signal processing.

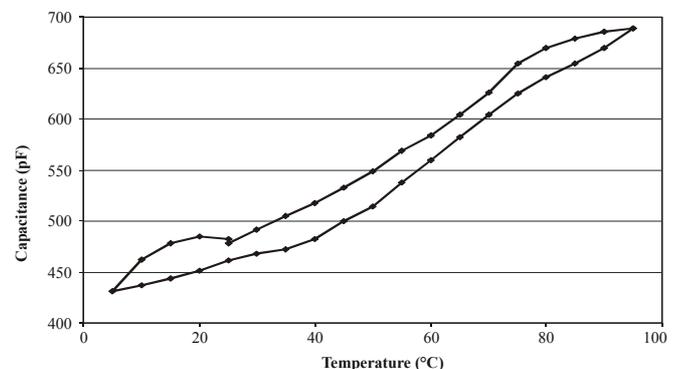

Fig. 4. The measured temperature–capacitance characteristics at constant 35% RH.

The resistive heating is divided to eight segments (see Fig. 1.) to ease the measurements and improve the yield in the development phase. Preliminary investigations show that it is possible to elevate the temperature above 80 °C at the center of the capacitive structure. Although the first results are promising it is necessary to optimize the parameters.

To decrease the power consumption and to improve the response time of the heating additional processing steps are planned to reduce the heated mass of silicon. Using backside anisotropic etching for bulk micromachining it would be possible to form a MEMS membrane below the capacitance and the surrounding heating elements.

*D.    Surface investigations: SEM, AFM*

For the direct observation of the alumina layer, SEM and AFM imaging were used. SEM images were taken by the aid of a LEO 1540XB CrossBeam® system. In Fig. 6. the mesopores can be seen clearly, the average mesopore size is about 9 nm. By enhancing the contrast of the images, populations of smaller pores were also found. We assume that they might be micropores with the radius less than 1 nm.

AFM measurements were performed to assess the roughness of the surface. The radius of the AFM tip was 20 nm and the measurements were made in contact mode. The biggest vertical deflection of the cantilever was 10 nm. According to the results of the AFM investigations, the alumina surface can be considered rather smooth.

*E.    Inner structure: Small Angle X-Ray Spectroscopy*

In order to obtain more precise information of the surface and also the inner structure of the alumina layer, Glazing Incidence Small Angle X-ray Scattering (GISAXS) experiments were carried out. A long strip of the surface was illuminated this way and scattering was generated by the inhomogeneities (electron density fluctuations) in the surface. We assumed that in case of anodized alumina the only inhomogeneities might be the pores.

A strip of absorbing lead was placed in front of the detector to eliminate both the direct beam and the specular reflections from the surface.

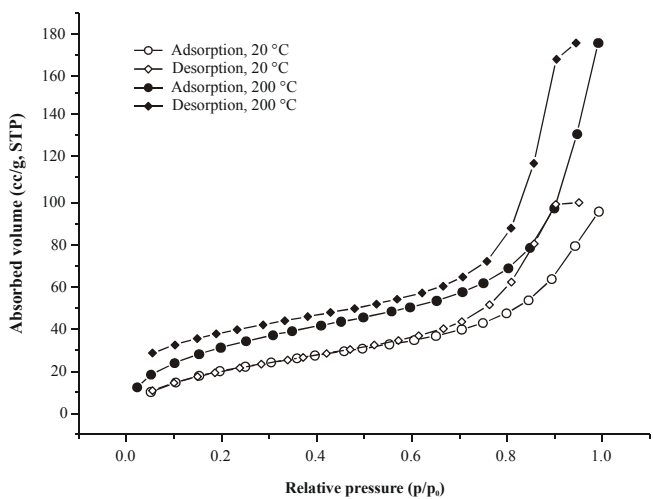

Fig. 5. Water-vapor adsorption isotherms on alumina powder.

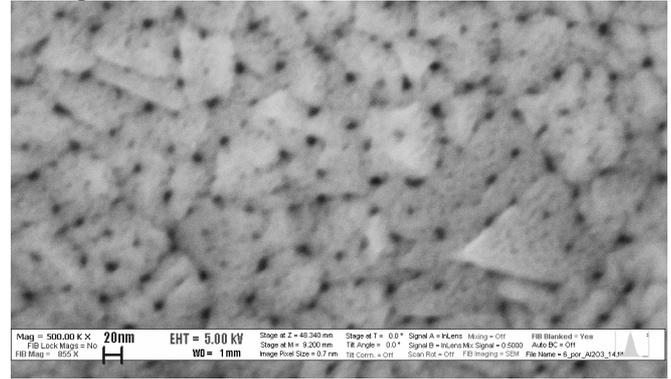

Fig 6. SEM image of the porous layer on the test-structure.

The GISAXS spectra (with the background subtracted) were measured at 3 different sectors having angles with respect to the horizontal plane: 22.5°-37.5°, 67.5°-82.5° and 127.5-142.5°.

A Lorentzian curve was fitted to the results:

$$I(q) = \frac{I(0)}{1 + r^2 q^2} \quad (5)$$

where $r$ is the average radius of the "pores" and $q$ is the wave vector. This results r=8.6 Å, i.e., a width of about 17 Å=1.7 nm.

Due to the low signal-to-noise ratio of the measurements, these results should be considered only semi-quantitative, still it is proven that the average pore sizes are in the microporous range, i.e. below 2 nm, and the system is fairly isotropic. This average pore size proved our estimation about the presence of micropores in the alumina layer, not clearly visible by SEM.

V.    CONCLUSIONS

The new AAO-based sensors are proven to be highly sensitive for ambient RH changes. The average sensitivity was approx. 15 pF/RH%, which is much higher than 0.2-0.5 pF/RH% found in the literature. The reason for the sensitivity improvement might be the increased average pore size we expect from the altered electrolyte concentration. Forthcoming SEM surface and cross-section investigations could prove this assumption.

The shape of characteristics is not linear, hysteresis and temperature-dependence is present due to physical laws and chemisorption of water molecules.

To overcome the parasitic temperature-dependence of the AAO RH sensors electronic signal processing or the continuous operation of the integrated heating could be used.

The processing of sensors is built on steps used in post-CMOS MEMS processing and has an additional electro-chemical oxidation step to prepare the porous thin film alumina sensing layer. This way the integration of read-out circuitry and a transceiver would be possible.

The integrated microheater for the conditioning of the sensing layer is working, but its parameters should be optimized. Its efficiency and response time could be significantly improved by reducing the heated mass with anisotropic backside etching, which would create a bulk micromachined MEMS membrane.

ACKNOWLEDGMENT

The authors wish to explain their gratitude to Dr. Krisztina László for the BET sorption measurements and the GISAXS opportunity. We are thankful to Dr. Éva Kiss for the wetting measurements, Dr. Attila Tóth for help in microscopic investigations and Brian Moffat for help in sensor processing at HWU.

The present research is a part of the development of a smart sensor and read-out circuitry for environment monitoring by employing new materials and semiconductor technology. Present research is partly funded by the Network of Excellence PATENT DfMM in FP6 of the European Union.